\documentstyle[epsfig]{aa}

\def\rs{\rm s}
\def\rs1{\rm s^{-1}}

\def\rcm{\rm cm}
\def\rcm2{\rm cm^{-2}}

\def\c2nor{\chi^2}

\begin{document}

\title{\emph{Beppo}SAX/PDS serendipitous detections at high galactic latitudes}

\author{R.~Landi\inst{1,2}
\and A.~Malizia\inst{1}
\and L.~Bassani\inst{1}
}

\offprints{R.~Landi: landi@bo.iasf.cnr.it}

\institute{INAF -- Istituto di Astrofisica Spaziale e Fisica Cosmica, Sezione di
Bologna, Via Gobetti 101, 40129 Bologna, Italy
\and
Dipartimento di Fisica, Universit\`a degli Studi di Bologna, Viale C. 
Berti Pichat 6/2, 40127 Bologna, Italy
}

\date{Received; Accepted}

\markboth{}{}

\abstract{At a flux limit of $\sim$$10^{-11}$ erg cm$^{-2}$ s$^{-1}$ in the
20--100 keV band, the PDS instrument on--board \emph{Beppo}SAX offers 
the opportunity to study the extragalactic sky with an unprecedented 
sensitivity. In this work we report on the results of a search 
in the \emph{Beppo}SAX archive for serendipitous high energy sources at
high galactic latitudes ($\vert b \vert\geq13^{\circ}$).
We have defined a set of twelve regions in which
the PDS/MECS cross--calibration constant is higher than the nominal value.
We attribute this mismatch to the presence of a serendipitous 
source in the PDS field of view.
In four cases the likely high energy emitter is also present in the MECS
field of view.
In these cases, we have performed a broad band spectral analysis 
(1.5--100 keV) so as to understand the source spectral
behaviour and compare it with previous \emph{Beppo}SAX observations 
when available.
In eight cases the identification of the source likely to provide the PDS spectrum
is based on indirect evidence (extrapolation to lower energies and/or comparison
to previous observations).
This approach leads to the discovery of
six new hard X--ray emitting objects (PKS 2356--611, 2MASX J14585116--1652223, NGC 1566,
NGC 7319, PKS 0101--649 and ESO 025--G002) and to the presentation the PDS spectrum of NGC 3227 
for the first time.
In the remaining five cases we provide extra \emph{Beppo}SAX observations that can be
compared with measurements which are already published and/or in the archive.
}
\authorrunning{Landi et al.}
\titlerunning{\emph{Beppo}SAX/PDS serendipitous detections at high galactic 
latitudes}{}

\maketitle

\keywords{X--rays: general; active galaxies; high energy emission -- method: 
data analysis 
}

\section{Introduction}
The hard X--ray sky is still poorly explored at high galactic latitudes and the only 
truly all sky survey
performed so far above $\sim$15 keV dates back to the 1980's
(\cite{Levine84}). This pioneering work was performed with the \emph{HEAO 1} A4 
experiment and detected $\sim$70 sources in the 13--80 keV band down to a flux limit of 
$2-3\times10^{-10}$ erg cm$^{-2}$ s$^{-1}$ with an angular resolution of 
$\sim$3$^{\circ}$; of these sources sixteen were located at 
high galactic latitudes ($\vert b \vert \geq13^{\circ}$) and only seven
were of extragalactic nature.

A step forward will be provided by the imager on--board \emph{INTEGRAL} which
is surveying a great fraction of the sky with a sensitivity better than a few
mCrab \footnote{For a Crab--like spectrum 1 mCrab corresponds to $2\times10^{-11}$ erg cm$^{-2}$ 
s$^{-1}$ in 
the 20--100 keV energy band} in the 20--100 keV energy range and an angular resolution of a 
few arcmin (Bassani et al. 2004, Bird et al. 2004); this exploratory work will
be followed by the \emph{SWIFT} mission which is expected to survey the hard X--ray
sky with a sensitivity of
$\sim$1 mCrab at high galactic latitudes (\cite{Gehrels04}). 
In the meantime, the \emph{Beppo}SAX archive can be used to probe the extragalactic
X--ray sky in the same 20--100 keV energy band.
Pointed observations of \emph{Beppo}SAX/PDS
have unveiled many hard X--ray emitting AGN and provided the best yet 
spectroscopy
of this type of objects above 10 keV. However, observations were sometimes limited by the
lack of imaging capability: contaminating sources were found inside the target field
of view as well as in the offset fields used for background measurements. These data
have often been neglected by the original observers although they provide another
powerful tool with which to study the extragalactic sky above 10 keV: new sources can be found
and spectroscopically measured, while known objects can be re--observed and compared
to previous measurements in a search for variability, which is a dimension
so far poorly explored. Furthermore, the argument put forward by 
Fabian (2001)\nocite{Fabian01} that two of the three nearest AGN have very high 
column densities ($> 10^{24}$ cm$^{-2}$) indicates 
the need to constrain the statistics 
of highly absorbed AGN: increasing the number of such objects known is
therefore an important task.
In view of these facts, we have carried out a program to search systematically in the 
\emph{Beppo}SAX/PDS archive (which at the moment is being reanalysed using the \emph{XAS} 
software package) for observations which indicate the presence of a 
contaminating source
either in the pointed field of view (search mode 1) or in the offset fields used
for background measurements (search mode 2).
Results related to search mode 1 are presented in this work, while a future work 
will be devoted to results found in search mode 2.

The paper is organized as follows: in Sec. 2 the selected sample is described, 
in Sec. 3 the observations, data 
reduction techniques and analysis procedures are presented, 
while Sec. 4 is devoted to the discussion of the 
results. Finally, conclusions and future work are summarized in Sec. 5.

%
%
%
%
%
%
%
%
\begin{table*}[t]
\caption{The list of serendipitous PDS detections found in the
\emph{Beppo}SAX archive.}
\smallskip
\small
\label{tab1}
\begin{center}
\begin{tabular}{l c l c l l}
\hline
\hline
{\bf Pointed}& {\bf Obs. date} &{\bf Type}& {\bf Calib$^{(a)}$ }&
{\bf Contaminating}&{\bf Type} \\
{\bf Source}   &   &   &   & {\bf Source}    \\
\hline
\hline
IRAS 01025--6423 & 24/03/2001 &Seyfert 2 & 18--1324 & PKS 0101--649 & Quasar     \\
MKN 1073 &  15/02/1999  &Seyfert 2 & 485--2035 &  Perseus/NGC 1275 (?) & Cluster/Seyfert 2   \\
NGC 1553 & 16/01/1997 & Normal Galaxy & 14--46 &NGC 1566 & Seyfert 1 \\
         &  16/11/1997 &   &        &         &     \\
AD Leonis & 23/04/1997 & Flare Star & 152--230 & NGC 3227 & Seyfert 1  \\
   &  01/05/1999 &   &    &    &           \\
   &  08/05/1999 &   &    &    &           \\
   &  12/05/1999 &   &    &    &           \\
ON 325 &  23/12/1998 & Blazar & 11--57 &  MKN 766$^{(b)}$  & Seyfert 1  \\
NGC 5793 & 25/07/2001   &Seyfert 2  & 32--1384 & 2MASX J14585116--165$^{(b)}$ & Galaxy   \\
1E 1839.6+8002  & 16/10/2000 &Flare Star &  83--181 & 3C 390.3$^{(b)}$
& Seyfert 1  \\
    &  03/02/2001 &     &    &     &    \\
H1846--786 & 08/03/2001 &Seyfert 1 &2--4 &  ESO 025--G002 & Seyfert 1  \\
VW Cephei & 07/05/1998& Eclipsing Binary &  164--417 & 4C +74.26 & Quasar  \\
   &  07/10/1998 &   &    &    &           \\
NGC 7331 &  10/06/2000 &LINER  & 28--121 & NGC 7319 & Seyfert 2       \\
NGC 7552 & 23/12/1999& Galaxy & 150--309 & NGC 7582  & Seyfert 2      \\
SCG 2353--6101& 28/11/1996 & Cluster  & 8--28 &PKS 2356-611$^{(b)}$
& Seyfert 2  \\
\hline
\hline
\end{tabular}
\begin{list}{}{}
\item[$^{(a)}$] PDS/MECS cross--calibration constant.
\item[$^{(b)}$] Source located within the MECS field of view (see text).
\end{list}
\end{center}
\end{table*}
\section{Sample selection and contaminating source search}
Although \emph{Beppo}SAX was not designed to perform an X--ray survey,
a systematic search in the entire PDS data archive could allow the discovery of new sources,
down to a flux limit of $\sim$$10^{-11}$ erg cm$^{-2}$ s$^{-1}$.
In particular, the simultaneous monitoring of three $1^{\circ}.3$ sky regions
(target field + two offset fields) allowed the PDS to survey a substantial fraction of the
sky over the 15--100 keV energy range.

The PDS instrument has a hexagonal field of view of $1^{\circ}.3\times1^{\circ}.3$ FWHM
and no imaging capability; its positional uncertainty can be approximated by
an error box circle of $1^{\circ}.3$ in radius.
The MECS has instead a field of view of $30^{\prime}$
radius and so covers about 25$\%$ of the PDS area.
It is therefore likely that a serendipitous source can be undetected
by the MECS, but still observed by the PDS.

In this paper we concentrate on the study of those sources discovered in the pointed
fields of view or detected in search mode 1.
The search in this mode was performed in the following way: first, we visually inspected in the 
\emph{Beppo}SAX archive (www.asdc.asi.it) all observations performed above 
$13^{\circ}$ in galactic latitude and available to the public as of October 
2001 (635);
from this set of data we extracted 
a sample containing those sources which clearly show a mismatch between the MECS and PDS spectra 
in the standard archive analysis: this mismatch was taken as strong evidence for the presence 
of a contaminating source in the PDS field of view.
From this preliminary list we excluded all those sources that were likely to be Compton thick 
on the basis of various considerations (i.e. large iron line equivalent width, X--ray to [$O_{III}$] 
ratio 
less than 1, extreme absorption): in the spectrum of these type of sources a MECS/PDS 
spectral mismatch is simply due to a more complex spectral shape than used for
the quick look analysis.
Then, we performed a cut in the signal to noise ratio,
accepting a source only if it had at least 3$\sigma$ detection in the PDS. We then confirmed with 
our own analysis the MECS/PDS mismatch by fitting the MECS and PDS data with a 
simple model, generally a power law either unabsorbed or absorbed: only 
when the
cross--calibration constant between these two instruments was significantly   
outside the nominal range of 0.75--0.95 (\cite{Fiore99a}), i.e. greater than 2, was a source
maintained. Overall, 
this analysis has provided a sample of twelve regions which are discussed in the present paper.
Although this sample is not complete (not all observations performed by \emph{Beppo}SAX were 
screened) nevertheless this search gives 
an idea of the extragalactic sources that populate the hard X--ray sky; in this sense the sources 
serendipitously found in this work can be ``loosely''
taken as representative of the AGN population in the 20--100 keV band.

An important step in the search described above, is related to the reduction of the PDS spectra 
which were extracted using the \emph{XAS} v2.1
package (\cite{Chiapp97}): this provides slightly different results than 
\emph{SAXDAS}, i.e. the package usually used to perform the archive analysis.
A preliminary comparison between PDS spectra extracted by means of these two software 
packages is under way\footnote{On~behalf~of~the~PDS~group,~see
ftp://ftp.tesre.bo.cnr.it in the directory
/pub/sax/doc/software\_docs/xas\_vs\_saxdas.ps.} and will be presented in a future work.
The analysis performed on sources  
of different intensity show that the spectral parameters do not change 
when computed with the different packages, but the associated errors are smaller 
when using \emph{XAS}. It is also important to underline that a significant  
improvement in the signal to noise ratio is obtained by means of \emph{XAS}.
Furthermore, the \emph{XAS} package allows a more reliable check of the 
background fields by taking advantage 
of the rocking technique (\cite{Fro97b}); this is rather important when the source is
faint in the PDS as is often our case. 
When one collimator is pointing ON source, the 
other collimator is pointing in one of the two OFF positions.
The standard stay time in each position of either collimator was 96 s. At each cycle
the two collimators were swapped: the one pointing to the source was moved to monitor
the background and vice versa.
In this way we can obtain, in addition to the target observation, two 
independently accumulated spectra of the two +OFF and --OFF fields which are offset by
$210^{\prime}$ with respect to the main pointing; these offsets  are used as background in the
computation of the spectrum of the target source. 
The comparison between the spectra of these two offset fields and, in particular, 
the difference between the +OFF and --OFF spectra in count rate, is a good diagnostic tool to 
investigate the presence of contamination. 

If no contamination is present in either of the offset fields, we expect the difference 
of their count rates to be compatible with zero; on the contrary,  
a positive excess of counts in the difference indicates the presence of 
contamination in the +OFF field; vice versa a negative count rate provides evidence for 
contamination in the --OFF field.
   
Applying this method to our sample, we found the presence of a contaminating
source in the +OFF field of the first observation of AD Leonis (excess: $0.189\pm 0.070$ counts
s$^{-1}$, 2.7$\sigma$)
and in part of the first observation of VW Cephei (excess: $0.139\pm 0.043$ counts
s$^{-1}$, 3.2$\sigma$), while in part of the 
observations of MKN 1073 and NGC 7552 the --OFF fields show an excess of $0.159\pm0.060$ counts 
s$^{-1}$ (2.7$\sigma$) and $0.203\pm0.0.056$ counts s$^{-1}$ (3.6$\sigma$), respectively.
To give an idea of the effect of contamination in an offset field on the
PDS data, we have considered two extreme cases: a $\sim$3$\sigma$ excess 
reduces the source count rate of $\sim$5$\%$, while a 10$\sigma$ contamination provides a
reduction at around 15$\%$.

In any case, in order to extract the 
uncontaminated source spectra, we excluded the contaminated fields
and considered only the uncontaminated one in the computation of the background
for these particular sources.

After assembling the sample, we search for the likely contaminating sources, adopting the
following strategy.
First, we considered all the sources present in the MECS 
image in addition to the target source and checked
their consistency with the PDS data, i.e. if the cross--calibration constant
fell within the nominal interval.
If no sources in the MECS field of view matched the PDS data, then,
we searched for likely high energy emitters located
inside the PDS field of view but not observed by the MECS given the 
significantly different fields of view.
In particular, when adopting this second
approach, we focused our search on bright sources in the 2--10 keV
band as these are the most likely to contaminate the PDS observation and are 
expected to appear in the HEASARC X--ray archives.
In Table~\ref{tab1} we list the 
twelve cases we found, reporting in each case the \emph{Beppo}SAX
observation target with its relative observation date and object type, the PDS/MECS 
cross--calibration 
constant obtained by fitting the data with a simple power law and finally the name
and type of the contaminating source found.
It is evident from the Table that the cross--calibration
constant is always outside the nominal range,
confirming the presence of one or more contaminating objects in the PDS field of view.
It is worth noting that extrapolation of a more complex
model from the MECS to the PDS energy range, still provides in all cases a high cross calibration 
constant.
We found only four fields where the contaminating source is so close 
($\sim$$25^{\prime}$) to the\emph{Beppo}SAX 
target that it is also detected by the MECS instrument.
%
%
%
%
\section{Observations and data analysis}
In this work we made use of data from three of the \emph{Narrow Field
Instruments}~(NFIs)~on--board~the~Italian--Dutch~satellite~\emph{Beppo}SAX~(\cite{Boella97a}):~
the~Low Energy Concentrator Spectrometer (LECS, 0.1--10 keV, \cite{Parmar97}), 
the Medium Energy Concentrator
Spectrometer (MECS, 1.3--10 keV, \cite{Boella97b}),
and the Phoswich Detection System (PDS, 15--300 keV, \cite{Fro97b}).

For all sources, the LECS and MECS spectra were 
downloaded from the \emph{Beppo}SAX archive; the analysis of the on--axis source 
was standard.
For all the off--axis sources detected in the MECS field of view, contaminating or not,
the MECS spectra were extracted from 
a region centered on the source and having a radius chosen according to the  
criteria suggested by Fiore et al. (1999a)\nocite{Fiore99a}.
For these sources, because of the lack of the appropriate ancillary
response files, the LECS spectral analysis could not be performed.
The background subtraction for the on--axis sources was performed using
blank sky spectra extracted from the same region of the source, while for the 
off--axis sources we used a local background spectrum (extracted from a 
region with a radius equal to the source extraction radius) 
to account for possible contaminating effects inside the MECS field of view.

For the PDS data reduction, source visibility windows were selected following the criteria
of no Earth occultation and high voltage stability during the
exposure. In addition, the observations closest to the South
Atlantic Anomaly were discarded from the analysis.

The LECS and MECS spectra were rebinned in order to sample the energy 
resolution of the detectors with an accuracy proportional 
to the count rate. The PDS data were instead rebinned so as to have 
logarithmically equal energy intervals. The data rebinning also
required that at least 20 counts was in each bin so that the
$\chi^{2}$ statistic could reliably be used. 

The energy bands used for spectral fitting were limited
to those where the response functions are well known, i.e. 0.1--4.5 keV,
1.5--10.5 keV, and 15--100 keV, for the LECS MECS, and PDS,
respectively. 

The spectral analysis was performed using the {\sc XSPEC v11.2.0}
software package  
(\cite{Arnaud96}) and the instrument response matrices released
by the \emph{Beppo}SAX ASI Science Data Center.
For the off--axis sources, we used the appropriate MECS ancillary response 
files to correct for the effects of vignetting due to the mirrors.
In addition, we introduced a flux correction factor in the PDS band
for each contaminating source in order to estimate the real 
flux at the source. This correction is simply a function of the distance of the source 
from the main target and is related to the PDS response: the reduction in  
sensitivity is of a factor of $\sim$2 at $38^{\prime}$ from the main pointing coordinates
while at $78 ^{\prime}$ the response is zero (\cite{Fro97a}). 

Normalization constants have also been introduced to allow for known differences 
in the absolute cross--calibration between the detectors. The LECS/MECS 
cross--calibration constant was allowed to lie within the nominal range 0.7--1.0
(\cite{Fiore99a}), while the PDS/MECS cross--calibration was left free
to vary in order to search for the presence of contaminating sources
and then to estimate their broad band spectrum. 
The absorption of X--rays due to our galaxy in the 
direction of each object (\cite{Diklog90}) is added in all models of 
the spectral analysis.
All quoted errors correspond to $90\%$ 
confidence interval for one interesting parameter ($\Delta\chi^{2}=2.71$).

The main aim of this work is to estimate the PDS spectrum of each serendipitous 
source found.
The method we used to achieve this was to fit simultaneously the LECS
(not always available) 
and MECS data of the target source to determine the best--fit model to these data; 
afterwards, we performed an extrapolation of this best--fit model to the
PDS energy band in order to estimate the contribution
of the target source to the total PDS flux.
Hence, by subtracting from the total PDS spectrum the contribution due to
the on--axis source, we were able to estimate the high energy spectrum of the 
serendipitous/contaminating object.

Four sources in our sample (NGC 1553, 1E 1839.6+8002, VW Cephei 
and AD Leonis), have been observed by \emph{Beppo}SAX at different epochs
(see Table~\ref{tab1}).
As a first step, our approach was to analyse each single pointing, for each source,
individually.
Although in some cases we found evidence for flux variability in the 
target and/or the serendipitous sources, the spectral parameters were always consistent
within the respective uncertainties; therefore,
we decided to sum, in all cases, all available pointings to improve
the statistical quality of the data, reporting only the averaged spectral
parameters.
%
%
%
%
%
%
%
%
\section{Results}
In the following, we briefly describe the main results 
obtained for each field. We discuss separately the
four fields in which the contaminating object is detected in the MECS instrument
because for these objects we can perform a broad band spectral analysis (1.5--100 keV), while in
the remaining eight cases we can only present and discuss the PDS spectrum.
\subsection{{\bf Sources detected in the MECS field of view}}
In this section, we describe the four fields where the object likely to be responsible 
for the PDS spectrum is also observed by the MECS instrument.
In Table~\ref{cont1} (and also in Table~\ref{cont2} afterwards) we report for each of these 
fields, all sources detected by the MECS,
their fluxes extrapolated from the MECS to the
PDS energy range (15--100 keV) assuming their best fit model, their relative 
contribution (in $\%$) to the PDS flux, their distance from the target 
source and the relative flux correction factor, $R$. 
The observed flux is divided by $R$ to reconstruct the flux at the source.
It is evident from the values reported in column 4 (Contribution) which 
source is in each case responsible for most of the PDS flux.
In Table~\ref{tab2} we list for each of these likely associations, the 
MECS and PDS exposures and count rates, the 2--10 keV flux,
the observed high energy flux (20--100 keV) and 
the effective flux (20--100 keV) at the source.
For these four sources we jointly fitted the MECS and PDS data to
provide a broad band (20--100 keV) spectrum.
In Table~\ref{tab3} we summarize the best--fit parameters and also
report the range of the cross--calibration constant PDS/MECS (column 6).
As is evident from the values of column 6, the normalization constants
obtained are always 
compatible with the suggested value of 0.75--0.95. 
This is a further indication that these sources
are indeed  responsible for the high energy emission detected
by the PDS instrument.
%
%
%
%
%
%
%
%
%
%
%
%
\begin{table*}[t]
\caption{Contribution to the PDS flux of the sources detected in the MECS field of view.}
\smallskip
\small
\label{cont1}
\begin{center}
\begin{tabular}{l l c c c l}
\hline
\hline
{\bf Field} & {\bf Source}  &  {\bf Extrapolated flux$^{(a)}$}&
{\bf Contribution} &  {\bf Offset$^{(b)}$}& {\bf R$^{(c)}$} \\
     &  &   [20--100] keV   &  ($\%$)  &  (arcmin)  & ($\%$)  \\
\hline
\hline
ON 325 & ON 325   & 1.2   & 10 &  -- & --     \\
   &  1AXG J121854+2957 & 0.56  &  3  & 16.2 & 78     \\
    & MKN 766 & 17.5 &  87   & 19.7 &73    \\
 NGC 5793 & NGC 5793   & 0.072 & 1 & --  & --    \\
   & NGC 5796 & 0.17 & 2  &  4.2  & 92      \\
   & 2MASX J1458--165  & 7.72 &   95 &  13.4  & 81   \\
1E 1839.6+8002 & 1E 1839.6+8002 &   0.017 & $<1$ & -- &  --     \\
   & 3C 390.3   &  47.8 & 98 &  25.3 &  65  \\
SCG 2353--6101 & SCG 2353--6101 &   0.03  & $<1$ & -- & --  \\
    & PKS 2356--611 & 25.2 & 99 & 16.1 &  78 \\
\hline
\hline
\end{tabular}
\begin{list}{}{}
\item[$^{(a)}$] In units of 10$^{-12}$ erg cm$^{-2}$ s$^{-1}$.
\item[$^{(b)}$] Distance between the pointing target and the off--axis sources.
\item[$^{(c)}$] Source flux correction factor due to the PDS response to source an off--axis 
source.
\end{list}
\end{center}
\end{table*}
\subsubsection{The ON 325 Field}
Within the MECS field of view (see Fig. 1) of this pointing we find two
other sources besides the target. The first, located at $\alpha(2000)=12^{h}18^{m}54^{s}.4$ and
$\delta(2000)=+29^{\circ} 58^{\prime} 10^{\prime\prime}.5$, is identified
with the Seyfert 1.9 galaxy 1AXG J121854+2957, which belongs to the
\emph{ASCA} Medium Sensitivity Survey (AMSS) (\cite{Ueda01}) and to the \emph{Beppo}SAX
High Energy Large Area Survey (HELLAS) (\cite{Fiore99b}). The second is
located at
$\alpha(2000)=12^{h}18^{m}25^{s}.2$ and $\delta(2000)=+29^{\circ} 48^{\prime}
48^{\prime\prime}.9$, $\sim$$20^{\prime}$ from ON 325
and corresponds to the Seyfert 1 galaxy MKN 766.

Following Perri et al. (2003)\nocite{Perri03}, the LECS and MECS data of the blazar
ON 325 are best fitted with a broken power law, for which we obtain
a $\chi^{2}/\nu=32.4/25$ and a 2--10 keV flux of $\sim$$8\times10^{-13}$
erg cm$^{-2}$ s$^{-1}$.
The two power laws intersect at 
$E_{B}=4.12^{+1.10}_{-0.83}$ keV and have photon indices
$\Gamma_{1}=2.41^{+0.18}_{-0.27}$ and $\Gamma_{2}=0.44^{+0.79}_{-0.85}$
in agreement with Perri et al. (2003)\nocite{Perri03}.
Adopting this model, we estimate a  contribution of ON 325
to the PDS data of the order of 10$\%$ (see Table~\ref{cont1}).
%
%
%
%
%
\begin{figure}[t]
\psfig{file=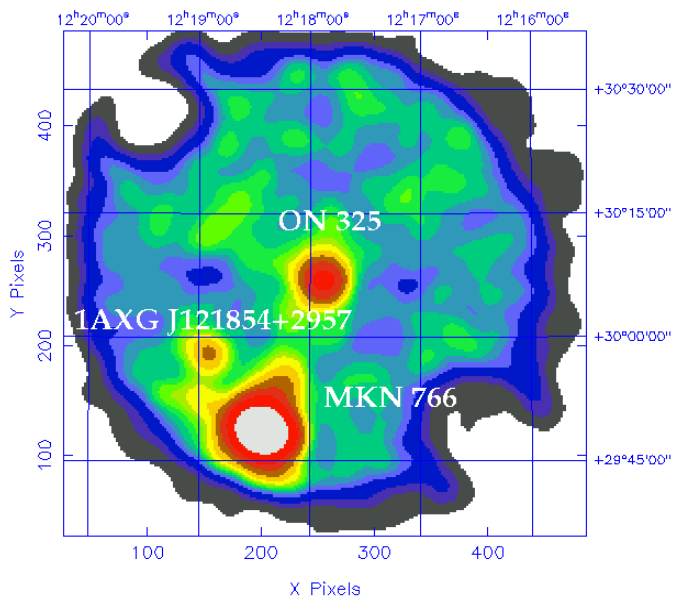,width=9.2cm,height=8cm,angle=0}
\caption{MECS image of the sky region surrounding ON 325.}
\end{figure}

The MECS spectra of 1AXG J121854+2957 is instead well fitted ($\chi^{2}/\nu=10.8/14$)
with an absorbed power law having  a photon index $\Gamma$ $\sim$2.3,
a column density N$_{\rm H}<12\times10^{22}$ cm$^{-2}$ and a 2--10 keV flux
of $8.5\times10^{-13}$ erg cm$^{-2}$ s$^{-1}$.
These values are in agreement with those found by Loaring et al. (2003)\nocite{Loaring03}
during a \emph{XMM--Newton} observation of the source. Extrapolation of this spectrum to the PDS band 
indicates 
a small contribution to the high energy flux, while a simultaneous fit to the MECS/PDS data provides 
a PDS/MECS cross--calibration constant in the range 2--30, i.e. well outside the nominal interval. 
A good fit ($\chi^{2}/\nu=30/34$) of the MECS data of MKN 766 is also described by a simple power law
with a photon index $\Gamma=1.92\pm0.13$ and a 2--10 keV flux of $1.7\times10^{-11}$ erg cm$^{-2}$
s$^{-1}$.
The high contribution to the PDS flux (see Table~\ref{cont1}) found with this model 
indicates that MKN 766 is the best candidate to account for the PDS flux.
In fact, the fit of the MECS spectrum of MKN 766 with the PDS data  gives
a PDS/MECS cross--calibration constant well inside the nominal interval.
The 2--10 keV flux (see Table~\ref{tab2}), found during this pointing, is
slightly lower than the value of $2.05\times10^{-11}$
erg cm$^{-2}$ s$^{-1}$ found by Matt et al. (2000a)\nocite{Matt00a} during
a previous (May 1997) \emph{Beppo}SAX dedicated observation.
We find no evidence for an iron line at around 6.4 keV (see Fig. 2), consistent
with the fact that this feature seems to be present only
when MKN 766 is in a high state (\cite{Leighly96}).
\begin{figure}[]
\psfig{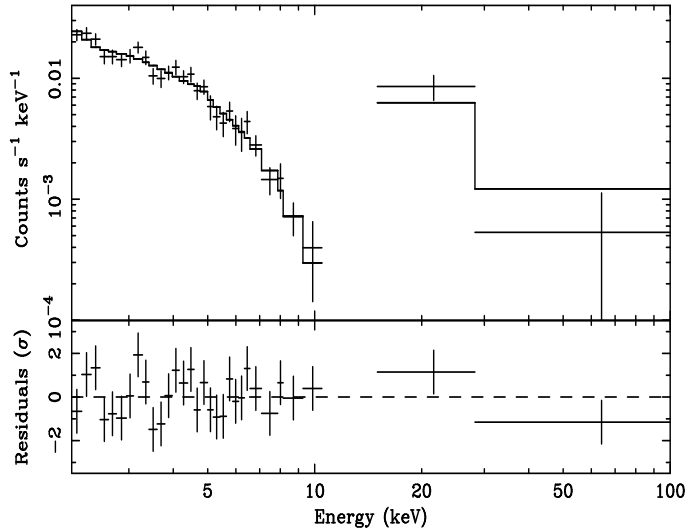}
\caption{Broad band spectrum of MKN 766 fitted with a simple power law (upper panel)
and residuals to this model in units of $\sigma$ (lower panel).
The data are graphically rebinned.}
\end{figure}

To confirm if MKN 766 is really the contaminating source in the ON 325 field,
we also reanalysed the PDS data
of the two pointings of MKN 766 performed by \emph{Beppo}SAX on
May 1997 and 2001.
In the first observation the comparison between the +OFF and
--OFF field count rates shows an excess of $3\sigma$ in the --OFF field,
while in the second there is no evidence for contamination.

Searching in the HEASARC archive for possible objects which could contaminate
the --OFF field,
we found two sources belonging to the
\emph{ROSAT} Bright Source Catalog: a quasar (1RSX~J121320.3+270841)
and an unidentified object (1RSX~J121258.3+272653),
at a distance of $\sim$40$^{\prime}$ and $\sim$59$^{\prime}$ from the field centre.
Fitting the --OFF data with a simple power law, we obtain
$\Gamma \leq 2$ and a 20--100 keV flux of
$3.6 \times 10^{-11}$ erg cm$^{-2}$ s$^{-1}$.
Since both sources have similar count rates in \emph{ROSAT}, we can only state  
that either or both could be responsible for this excess.

In any case, this provides a good example of the capability of our analysis:
contamination in the offset fields can be found and identified, while at the same
time a more correct evaluation of the target spectrum is provided.

The PDS spectrum of MKN 766 in the May 1997 observation can be corrected
and reanalysed together with that of May 2001. In both cases the photon index
$\Gamma$ is $\sim$2, in agreement with that found in December 1998 (i.e. during
the ON 325 observation),
while the 20--100 keV flux is roughly $\sim$$3\times10^{-11}$ erg cm$^{-2}$ s$^{-1}$,
higher than during the ON 325 measurement; this finding indicates that the flux
variability always seen in this source at soft X--ray energies
(\cite{Matt00a} and references therein) extends to above $\sim$10 keV.
\begin{table*}[t]
\caption{MECS and PDS Observation Log.}
\smallskip
\small
\label{tab2}
\begin{center}
\begin{tabular}{l c c c c c c c}
\hline
\hline
{\bf Source}&\multicolumn{2}{c}{\bf MECS} &\multicolumn{2}{c}{\bf PDS}
& {\bf F$^{(a)}$}&{\bf F$^{(a)}_{Obs}$}
& {\bf F$^{(a)}_{Source}$} \\
  &   {\bf EXPO} & {\bf cts/s} & {\bf EXPO} & {\bf cts/s} &   &  &  \\
   & (ks)  &  & (ks)   &  & [2--10] keV &[20--100] keV &
 [20--100] keV   \\
\hline
\hline
MKN 766 & 31.4 & $0.0153\pm0.0011$ & 15.7 & $0.152\pm0.050$
&16.0 & 17.0 & 23.3 \\
2MASX J1458--165 & 78.7 &$0.0042\pm0.0003$& 38.5 &$0.086\pm0.029$&
0.93& 8.1 & 11.1 \\
3C 390.3$^{(b)}$& 157.3 & $0.078\pm 0.001$&80.0 & $0.447\pm0.002$
 & 23.0  & 22.0 & 43.0 \\
PKS 2356--611 & 23.4 &$ 0.029\pm0.001$ & 25.4 &$0.205\pm0.041$
& 13.0 & 25.5  & 32.7 \\
\hline
\hline
\end{tabular}
\begin{list}{}{}
\item[Note:]  Count rates are referred to the energy bands used in the
spectral analysis (see text).
\item[$^{(a)}$] In units of 10$^{-12}$ erg cm$^{-2}$ s$^{-1}$.
\item[$^{(b)}$] For this source the parameters of the average spectrum
are reported.
\end{list}
\end{center}
\end{table*}
\begin{table*}[]
\caption{MECS and PDS spectral analysis of the serendipitous sources
detected within the MECS field of view.}
\smallskip
\small
\label{tab3}
\begin{center}
\begin{tabular}{l c c c c c c}
\hline
\hline
{\bf Source}&{\bf ${\Gamma}$} &{\bf N$^{(a)}_{\rm H}$}&{\bf E$_{Line}$}&
{\bf EW}& {\bf Calib$^{(b)}$}& {\bf ${\chi}^{2}/\nu$} \\
    &   &     & (keV)   & (eV)    &  &    \\
\hline
\hline
MKN 766 & $1.93\pm0.10$ & -- & -- & -- & 0.5--1.3 &  36.4/36 \\
2MASX J1458--165 & $1.20^{+0.26}_{-0.23}$ & -- & -- & -- & 0.8--3.2 & 16.1/24 \\
3C 390.3$^{(c)}$& $1.72^{+0.04}_{-0.05}$ & $<0.08$  &  
$6.37^{+0.29}_{-0.25}$& $115^{+86}_{-84}$  & 0.89--1.22 & 16.5/22 \\
PKS 2356--611 &$1.70^{+0.18}_{-0.17}$ &$10.2^{+2.91}_{-2.31}$&$6.46^{+0.21}_{-0.17}$ &
$523^{+238}_{-246}$ & 0.75--0.95 & 48.5/41 \\
\hline
\hline
\end{tabular}
\begin{list}{}{}
\item[$^{(a)}$] In units of $10^{22}$ cm$^{-2}$.
\item[$^{(b)}$] PDS/MECS cross--calibration constant.
\item[$^{(c)}$] For this source the parameters of the average spectrum 
are reported. 
\end{list}
\end{center}
\end{table*}
\subsubsection{The NGC 5793 Field}
NGC 5793 has been detected at $\sim$$5\sigma$ level
in the MECS instrument, while it is insignificant ($<1\sigma$) in the LECS
energy range. 
Fitting the MECS data with a simple
power law of photon index $\Gamma=1.9$ we obtain a very low flux of 
$\sim$$7\times10^{-14}$ erg cm$^{-2}$ s$^{-1}$. This model,
extrapolated to the high energy band, gives a negligible
contribution to the PDS data (see Table~\ref{cont1}).
%
%
%
%
\begin{figure}[t]
\psfig{file=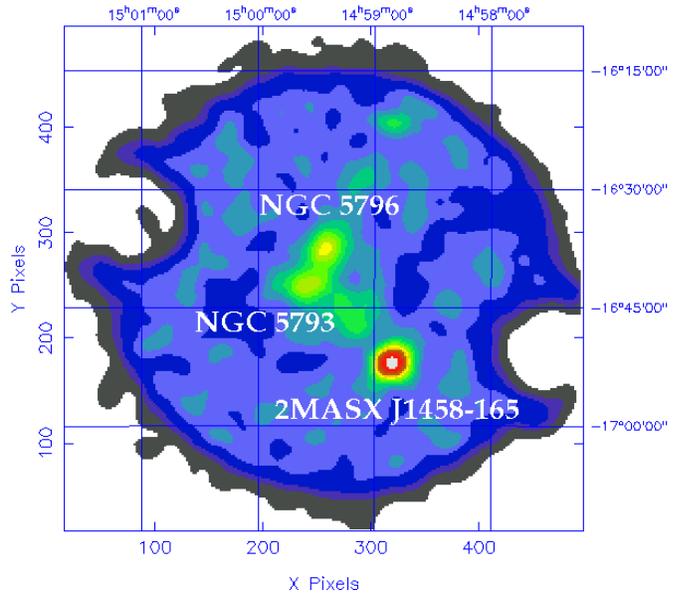,width=9.2cm,height=8cm,angle=0}
\caption{MECS image of the sky region surrounding NGC 5793.}
\end{figure}

Inspection of Fig. 3 indicates the presence of two extra
sources in the MECS field of view. The first located at $\alpha(2000)=
14^{h}58^{m}50^{s}.7$ and
$\delta(2000)=-16^{\circ} 51^{\prime} 50^{\prime\prime}$.7, corresponds
to a bright irregular spiral galaxy 
belonging to the 2 Micron All Sky Survey eXtended Source Catalog
(i.e. 2MASX J14585116--1652223). The second is located at $\alpha(2000)=
14^{h}59^{m}23^{s}.8$ and $\delta(2000)=-16^{\circ} 37^{\prime}
13^{\prime\prime}.4$ and is identified with the normal galaxy NGC 5796.

We extracted the MECS spectra of both sources and extrapolated them
to high energies. We find that the MECS spectrum of NGC 5796 
is well fitted ($\chi^{2}/\nu= 15/20$) with a simple power law
having a photon index $\Gamma=1.95^{+0.67}_{-0.56}$ and a 2--10 keV flux of
$1.5\times10^{-13}$ erg cm$^{-2}$ s$^{-1}$. The contribution of NGC 5796 to 
the PDS flux is only a few percent and the PDS/MECS cross--calibration constant 
is extremely high (20--103). 

The MECS best--fit model
in the case of 2MASX J14585116--1652223 is a simple power law with photon index 
$\Gamma=1.15^{+0.33}_{-0.18}$
and a 2--10 keV flux of $9.3\times10^{-13}$ erg cm$^{-2}$ s$^{-1}$ ($\chi^{2}/\nu= 15/20$).
The extrapolation of this model to high energies (see Table~\ref{cont1}) 
suggests that  2MASX J14585116--1652223 is likely to be the contaminating object.
The simultaneous fit of the MECS/PDS data (see Table~\ref{tab2} and Fig. 4) 
shows a flat spectrum 
($\Gamma=1.20^{+0.26}_{-0.23}$), a more suitable value (0.7--3.5) for the PDS/MECS 
cross--calibration constant and no evidence for extra absorption.
Fixing $\Gamma=1.9$ and adding absorption to the power law provides a similar fit
and an upper limit to the column density of $\sim$$4\times10^{22}$ cm$^{-2}$.
As displayed in Table~\ref{tab2}, the 20--100 keV flux turns out to be
of $1.11\times10^{-11}$ erg cm$^{-2}$ s$^{-1}$, one
order of magnitude higher than the 2--10 keV flux,
confirming that 2MASX J14585116--1652223 emits mostly in the high energy band.
Very little is known of this object except for its near infrared properties:
the source is fairly bright with total photometry of 14.4, 13.6 and 13 magnitudes
in the J, H and K bands respectively and an extent of $11^{\prime\prime}$; the optical
counterpart has magnitudes B $ = 16$ and R $=15$. 
No previous X--ray data are reported, including the lack of detection by \emph{ROSAT},
which would be consistent with the presence of strong absorption.
The R--K colour is $\sim$2, i.e. quite red which again point to an object with considerable 
extinction.
No detection in the radio band was found in the literature or in the HEASARC archive.
\begin{figure}[]
\psfig{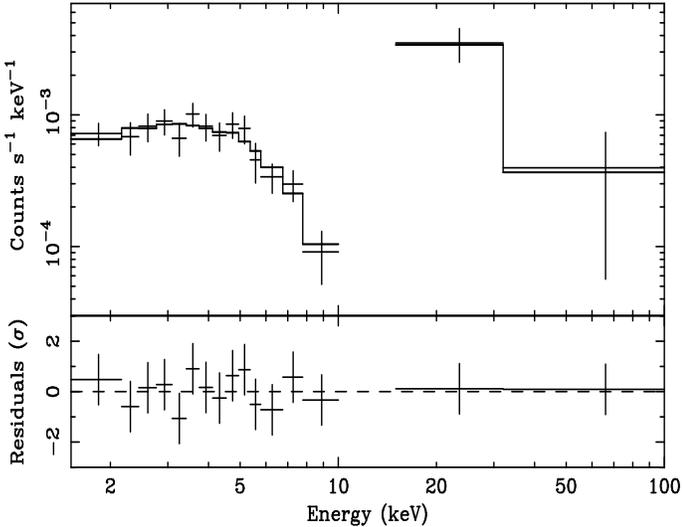}
\caption{Broad band spectrum of 2MASX J1458--165 fitted with a simple power law
(upper panel) and residuals to this model in units of $\sigma$ (lower panel). The data are 
graphically rebinned.}
\end{figure}
\subsubsection{The 1E 1839.6+8002 Field}
The M4 Ve star 1E 1839.6+8002 was observed by \emph{Beppo}SAX
twice (October 2000 and February 2001). This source was detected for the 
first time during the \emph{Einstein} observation
of the Broad Line Radio Galaxy 3C 390.3 in the 1980's.
We stress that a detailed study of the spectral behaviour
of this source is beyond the aim of this paper. Nevertheless,
following Pan et al. (1997)\nocite{Pan97}, the LECS and MECS spectra of both
observations are well described
by a thermal model ({\sc Raymond--Smith} model in XSPEC, \cite{Ray77}),
with a temperature $kT$ $\sim$1.8 keV
and a 2--10 keV flux of $9.5\times10^{-14}$ erg cm$^{-2}$ s$^{-1}$
(first observation) and $7.2\times10^{-14}$ erg cm$^{-2}$ s$^{-1}$
(second observation). 
%
%
%
%
%
\begin{figure}[t]
\psfig{file=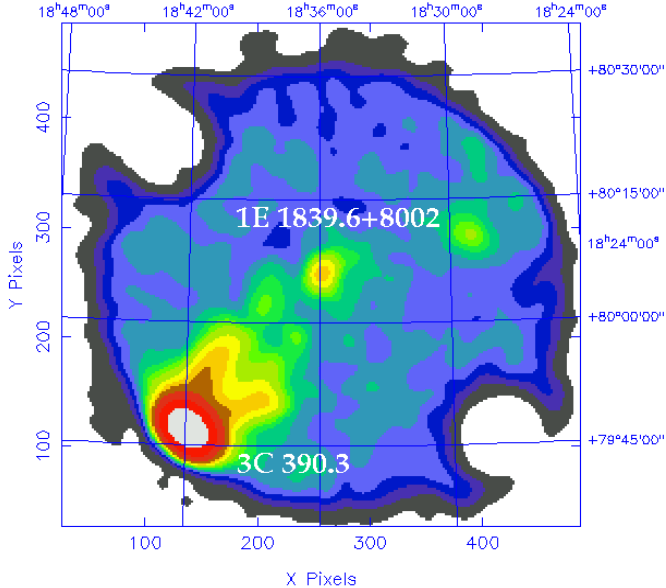,width=9.2cm,height=8cm,angle=0}
\caption{MECS image of the sky region surrounding 1E 1839.6+8002.}
\end{figure}
\begin{figure}[]
\psfig{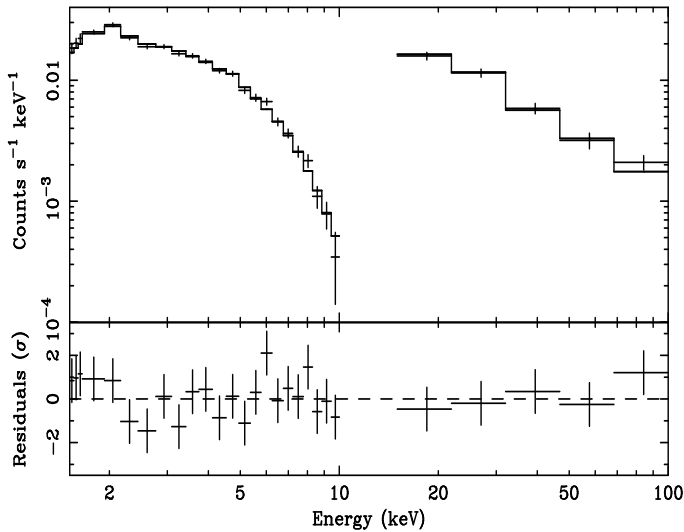}
\caption{Broad band spectrum of 3C 390.3 fitted with an absorbed power law
plus a narrow gaussian line
(upper panel) and residuals to this model in units of $\sigma$ (lower panel).}
\end{figure}

The source appears to be faint and
variable ($\sim$25$\%$) in the 2--10 keV range, with a negligible
contribution to the high energy emission in both observations (in 
Table~\ref{cont1} only the averaged contribution of the source is reported).

Within the MECS field of view (see Fig. 5)
of both measurements, a bright object located at
$\alpha(2000)=18^{h}41^{m}51^{s}.5$ and $\delta(2000)=+79^{\circ}47^{\prime}
01^{\prime\prime}.3$ is clearly visible: it corresponds to the well known radio galaxy 3C 390.3.
The \emph{Einstein} IPC data of 3C 390.3 revealed the presence of strong intrinsic
absorption (\cite{Krup90}), while
at higher energies \emph{Ginga} data showed the presence of an
iron line at 6.4 keV (\cite{Inda94}) and a reflection component
(\cite{Nandra94}). \emph{ASCA} data confirmed the presence of the iron K emission line
but could not constrain the reflection hump,
most probably because of the limited energy range (\cite{Eracleous96}; \cite{Leighly97}). 
3C 390.3 has also been detected by
\emph{OSSE} in the soft--$\gamma$ energy range, above 50 keV (\cite{Dermer95}).
All these results have been confirmed by Grandi et al.
(1999)\nocite{Grandi99}, on the basis of a dedicated \emph{Beppo}SAX observation performed
in January 1997.
The above authors also found a column density variability when
comparing \emph{Beppo}SAX results with previous measurements.

No evidence for variability is present in the MECS data of 3C 390.3 nor
in the PDS analysed in this work.
Hence, in order to improve the statistics, we summed the two data sets
together to perform our spectral analysis.
First, we analysed only the MECS data of 3C 390.3, finding as best--fit model
($\chi^{2}/\nu=13.5/17$) a power law with a slope of 1.7 plus
a narrow gaussian line having a centroid energy of 6.4 keV and an equivalent width 
of $EW= 115$ eV, compatible with
K$_{\alpha}$ line emission from neutral iron located at the source redshift.
The contribution of the 3C 390.3 flux to the total PDS flux
is significant (see Table~\ref{cont1})
and clearly indicates that this source is indeed responsible for the contamination.
 
The joint fit of the MECS data of this source with the PDS is also well
reproduced by a power law plus a narrow gaussian line 
($\chi^{2}/\nu=16.5/22$) having parameters similar to those found fitting only the MECS data. 
As for the column density, we can only provide an upper limit
of $\sim$$8 \times 10^{20}$ cm$^{-2}$, compatible with the values found and
reported by Grandi et al. (1999)\nocite{Grandi99}.
The broad band spectrum of the source is shown in Fig.~6.
The flux turns out to be $2.2\times10^{-11}$ erg cm$^{-2}$ s$^{-1}$
and $4.3\times10^{-11}$ erg cm$^{-2}$ s$^{-1}$ in the 2--10 and 20--100
keV band respectively, again in agreement with that found by 
Grandi et al. (1999)\nocite{Grandi99}. The PDS/MECS cross--calibration
constant turns out to be consistent (see Table~\ref{tab3}) with the suggested
values, confirming that 3C 390.3 is the serendipitous source contaminating the PDS 
emission during the observations of the 1E 1839.6+8002 field of view (see 
Table~\ref{cont1}).
There is no evidence for variability between the present (2000--2001)
\emph{Beppo}SAX observations
and that reported by Grandi et al. (1999)\nocite{Grandi99} in the 1997 observation.
\subsubsection{The SCG 2353--6101 Field}
In the case of the cluster of galaxies SCG 2353--6101 (Abell 4067) there is
another source in the
MECS field of view, PKS 2356--611, located at $\alpha(2000)=
23^{h}59^{m}04^{s}.1$
and $\delta (2000) = -60^{\circ} 55^{\prime} 01^{\prime\prime}.5$
(see Fig. 7).

PKS2356--611 ($z=0.096$, B $ =12.3$, R $ =12.8$, as reported in the USNO--B1 Catalog 
(\cite{Monet03})) is one of the strongest southern FR II sources, with a
total radio power $P_{\rm 1.4 GHz}$ $\sim$$10^{25.8}$ W Hz$^{-1}$
(\cite{Koek98}), while optically it shows strong high--excitation
narrow line emission with [O$_{III}$]$\lambda$5007/H$_{\beta}$ $>$ 10.
Lipovetsky et al. (1988)\nocite{Lipov88}
have cataloged this source as a Seyfert galaxy of type 2.
It is also listed in the 2 Micron All Sky Survey eXtended Source Catalog
as 2MASX J23590436--6054594 (with magnitudes J $=13.7$, H $ =12.7$, K $ =12.9$)
and was not detected by \emph{ROSAT}.

The LECS and MECS spectra of SCG 2353--6101 are well fitted
($\chi^{2}/\nu= 32/37$) with a thermal bremsstrahlung model, giving
a temperature of $4.74^{+1.70}_{-2.26}$ keV. The extrapolation of this model
to the PDS energy band provides negligible flux.
(see Table~\ref{cont1}). This, combined
with the high value (8--28) of the PDS/MECS cross--calibration, indicates the
presence of a high energy emitting source and PKS2356--611 is a good candidate.

As shown in Fig. 7, PKS2356--611 is located near one of the two $^{55}$Fe
calibration sources of the MECS instrument. In this case a careful choice of the 
background region is required to avoid contamination by 5.95 keV photons 
produced by the calibration source: this is done taking the background region as near as possible 
to the calibration sources itself.

Fitting the MECS data of PKS 2356--611 and the PDS points
with a single power law
plus intrinsic absorption provides an acceptable fit ($\chi^{2}/\nu= 60/43$),
a steep spectrum ($\Gamma=1.75$) and a column density of
$N_{\rm H}$ $\sim$$1\times10^{23}$ cm$^{-2}$ (Fig. 8).
Both spectral parameters are typical of
Seyfert 2 galaxies, and seem to confirm the source classification made by
Lipovetsky et al. (1988)\nocite{Lipov88}.

The cross--calibration constant in this case turns out to be lower (0.2--0.5) than
generally observed. This could be due to imperfect background correction in the MECS
because of the location of the source near the calibrator. In any case, the
spectral parameters do not change significantly when this constant is constrained to
vary within the nominal range of values.
%
%
%
%
%
\begin{figure}[t]
\psfig{file=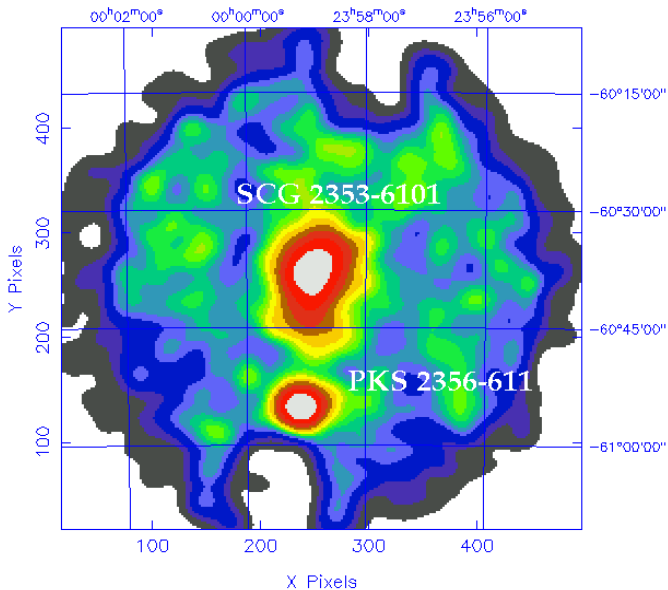,width=9.2cm,height=8cm,angle=0}
\caption{MECS image of the sky region surrounding SCG 2353--6101.}
\end{figure}
%
%
%
%
%
%
\begin{figure}[h]
\psfig{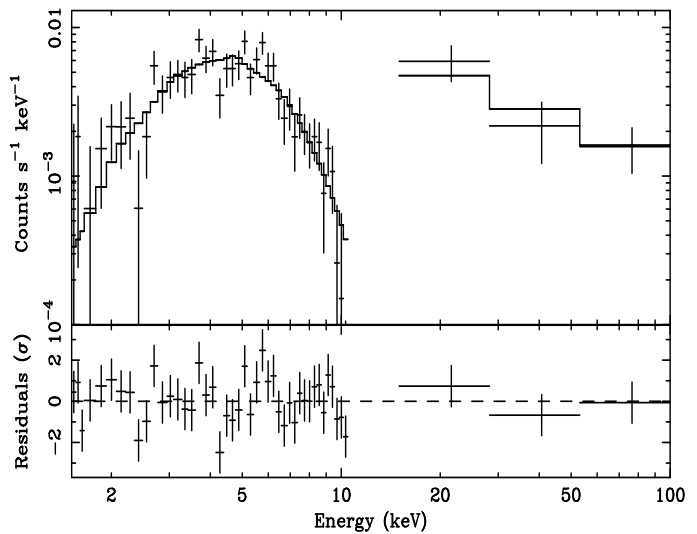}
\caption{Broad band spectrum of PKS 2356--611 fitted with an
absorbed power law (upper panel) and residuals to this model in units of
$\sigma$ (lower panel).}
\end{figure}

An inspection of the broad band spectrum of PKS 2356--611 (see Fig. 8)
suggests the possible presence of an excess
around $\sim$6 keV.
Although the addition of an extra component in the
form of a narrow gaussian line at around 6.4 keV with $EW$ $\sim$500 eV
provides an
improvement in the fit ($\Delta \chi^{2}= 11.5$ for two additional parameters),
its reality is questioned by the location of the source near the $^{55}$Fe calibrator:
residuals at line energy could still be due to a non perfect subtraction of the 
calibration line.
Also in this case the cross--calibration constant turns out to be
low ($<$ 0.5) and the $\chi^{2}$ values as well as the spectral parameters do not
change significantly if we vary the cross--calibration constant within the allowed range.
The absorption measured would produce an iron line of 100--200 eV
in transmission, while reflection in the torus would increase the $EW$
to about 400 eV (\cite{Turner97})
possibly indicating the presence of a reflection component.
However, the addition in the fit of this
component via the {\sc pexrav} model in XSPEC (\cite{Magd95})
is not required by the data and does not improve the cross--calibration constant, which
turns out to be very small ($< 0.4$).
A large $EW$ is also possible if the source is Compton thick; this type of object 
is generally
characterized by a low $F_{\rm X}$/$F_{\rm OIII}$ ratio (\cite{Bassani99}).
In PKS 2356--611 this ratio is $\sim$67, when $F_{\rm O{III}}$ is corrected for
the reddening in the galaxy using the observed value of $H_{\alpha}$/$H_{\beta}=4.32$
(\cite{Koek98}). This result suggests that PKS 2356--611 is a Compton thin Seyfert 2 galaxy
rather than a Compton thick one and so the large $EW$ is probably not due to strong absorption
($N_{\rm H} > 1.5\times10^{24}$ cm$^{-2}$).
%
%
%
%
%
%
%
\subsection{{\bf Sources not detected in the MECS field of view}}
In this section we describe the remaining eight fields where the source responsible
for the PDS spectrum is not observed by the MECS. 
In Table~\ref{cont2} we report for each of these fields, the source/s observed in the MECS,
their flux extrapolated to the 20--100 keV band and their contribution to the PDS flux.   
In Table~\ref{tab4} instead we list for each field, 
the PDS exposure and count rate as well as  
the photon index $\Gamma$ and observed 20--100 keV flux obtained assuming a simple power law fit. 
We also report the
effective flux at the source, estimated from that observed by applying the correction 
factor $R$.
\subsubsection{The IRAS 01025--6423 Field}
In this field only the target source, IRAS 01025--6423, a Seyfert 2 galaxy is 
barely detected by MECS at a $\sim$2$\sigma$ level, while the LECS data are not available.
Due to the low signal to noise ratio, we
only attempted to put an upper limit on the flux assuming a power law model with
photon index $\Gamma=2$. The source appears to be faint in the 2--10 keV
range with a flux $\leq 3.1\times10^{-14}$ erg cm$^{-2}$ s$^{-1}$
and its contribution to the PDS signal is obviously null (see Table~\ref{cont2}).
Searching in the HEASARC archive for possible contaminating sources
within the PDS field of view, we found two possible
candidates belonging to the \emph{ROSAT} Bright Source Catalog:
1RXS J011316.2--641142 (CPD--64 120), classified as a K1 Ve star,
and 1RXS J010333.7--643925 (PKS 0101--649), identified as a radio source, at distances
from the target source of $\sim$58$^{\prime}$ and $\sim$33$^{\prime}$, respectively.
PKS 0101--649, which is the closest to the target source, is
more likely to be the object
responsible for the high energy emission, as a star is not expected to emit
above 10 keV. PKS0101--649
belongs to the 2 Micron All Sky Survey eXtended object (i.e. 2MASX J01033376--6439079 in NED),
with total photometry of J $ =15$, H $ =14$ and K $ =13.4$ magnitudes; the USNO--A2 R 
magnitude is 16.5 (\cite{Monet99}) thus providing an R--K value of 3.1 which suggests a quite 
red object; at radio frequencies it is fairly bright with a 6 cm flux of $\sim$270 mJy.
\begin{table*}[t]
\caption{Contribution to the PDS flux of all the MECS sources in the eight fields in which the
contaminating source was not detected in the MECS field of view.}
\smallskip
\small
\label{cont2}
\begin{center}
\begin{tabular}{l l c c c l}
\hline
\hline
{\bf Field} & {\bf Source}  &  {\bf Extrapolated flux$^{(a)}$}&
{\bf Contribution} &  {\bf Offset$^{(b)}$}& {\bf R$^{(c)}$} \\
     &  &   [20--100] keV   &  ($\%$)  &  (arcmin)  & ($\%$)   \\
\hline
\hline
 IRAS 010125--6423 & IRAS 010125--6423   & 0.002 & $<1$ & --  & --    \\
 MKN 1073  & MKN 1073 & -- & --  &  --  & --      \\
NGC 1553 & NGC 1553  & 1.5  & 7  & -- & --  \\
Ad Leonis   & AD Leonis   &  0.0002 & $<1$ & -- &  --  \\
 H1846--786  & H1846--786  & 10 &   50 & --   & --   \\
VW Cephei & VW Cephei & 0.0004 & $<1$ & -- &  --     \\
NGC 7331    & NGC 7331 & 0.59 & 4 & -- &  -- \\
NGC 7552 & NGC 7552   & 0.12   & $<1$ &  -- & --     \\
   &  FRL 1041 & 1.78  &  4  & 18.9 & 76    \\
    & Sersic 159--03 & 0.003 &  $<1$   & 25.8 & 65    \\
\hline
\hline
\end{tabular}
\begin{list}{}{}
\item[$^{(a)}$] In units of 10$^{-12}$ erg cm$^{-2}$ s$^{-1}$.
\item[$^{(b)}$] Distance between the pointing target and the off--axis sources.
\item[$^{(c)}$] Source flux correction factor due to the PDS response to source not
on--axis.
\end{list}
\end{center}
\end{table*}

The PDS spectrum is characterized by a photon index
$\Gamma$ $\sim$1.5 (see Table~\ref{tab4}) compatible, within uncertainties,
with an AGN spectrum; the 20--100 keV flux is  $2.3\times10^{-11}$ erg
cm$^{-2}$ s$^{-1}$, making PKS 0101--649 a bright source at these energies.
If we extrapolate this power law to soft X--ray energies, we find a 0.1--2.4 keV flux
of $0.22\times10^{-11}$ erg cm$^{-2}$ s$^{-1}$ which is consistent with the \emph{ROSAT} value
of PKS 0101--649. Alternatively, the X--ray source spectrum could be steeper but highly absorbed 
even above 10 keV as observed in Compton thick objects (\cite{Matt00b}).
If we add absorption to the power law, fixing the photon index to 1.9, we find 
an upper limit to the column density $< 2 \times 10^{25}$ cm$^{-2}$.
Obviously, an optical spectrum would be highly desirable to provide more information and establish 
the true nature of this source.
\subsubsection{The MKN 1073 Field}
The Seyfert 2 galaxy MKN 1073 was not clearly detected by the LECS and
the MECS instruments. In both detections there is evidence for diffuse emission
probably associated with the Perseus cluster,
which is at $62^{\prime}$ from the centre of the MECS pointing.
The signal in the PDS is high ($>10\sigma$ level), with an exceptionally
steep spectrum ($\Gamma$ $\sim$3.3, see Table~\ref{tab4}).
We ascribe the PDS emission to the Perseus cluster, which has also been
targetted by \emph{Beppo}SAX on September 1996.
The spectral analysis of these PDS data also shows a very steep spectrum
($\Gamma=3.80^{+0.17}_{-0.18}$) and a 20--100 keV flux of
$\sim$$5\times10^{-11}$ erg cm$^{-2}$ s$^{-1}$, lower than
the flux seen during the MKN 1073 observation (see Table~\ref{tab4}). Since flux variability in the 
cluster
is unlikely, the change in the emission could be due to the Seyfert 2 galaxy NGC 1275,
often taken as responsible for the high energy ($>$ 20 keV) emission observed from this
region; comparison of data from high energy instruments over years shows a large
variation in the flux above 20 keV (\cite{Osako94}).
It is also possible that both cluster and AGN contribute to the high energy emission,
but it is difficult at present to disentangle one contribution from the other.
\subsubsection{The NGC 1553 Field}
\emph{Chandra} observations of the X--ray faint SO galaxy
NGC 1553 (\cite{Blant01})
has spatially and spectrally resolved the source of the X--ray emission.
A significant fraction of this ($\sim$70$\%$) is detected as diffuse flux
while the remainder is due to 49 objects.
The strongest source in the field is located
at the centre of NGC 1553 to within $0^{\prime\prime}$.5 and shows a hard
spectrum, typical of an AGN.
NGC 1553 was observed twice by \emph{Beppo}SAX (January 1997 and
November 1997). Trinchieri et al. (2000)\nocite{Trinc00}
performed the spectral analysis of the
LECS and MECS data finding as best--fit
a {\sc Raymond--Smith} model ($kT$ $\simeq$0.26 keV) for the soft component and a
thermal bremsstrahlung ($kT$ $\simeq$4.8 keV) for the hard component.
Our analysis of the LECS and MECS data from both observations is in agreement with
that reported by these authors;
the extrapolation of this model to high energies provides a very low contribution 
to the PDS flux (see Table~\ref{cont2}).
Searching in past X--ray mission archives, we
find that the Seyfert 1 galaxy NGC 1566 is the best candidate for the PDS emission as also 
suggested by Trinchieri et al. (2000)\nocite{Trinc00}.
This source is at a distance of $60^{\prime}$.2 from the target object
and is therefore just inside the PDS field of view.
NGC 1566 ($z= 0.005$, B $ =9.9$, R $ =9.6$) 
is a very bright X--ray source known since the \emph{HEAO 1} X--ray
Source Catalog (\cite{Wood84}).
The source belongs also to the 2 Micron All Sky Survey eXtended Source Catalog
(2MASX J04200041--5456161), with magnitudes J $ =7.8$, B $ =7.2$ and K $ =6.9$,
it is reported by \emph{ROSAT} in the Bright Source
Catalog and it is a radio source of $\sim$100 mJy at 6 cm.
The 2--10 keV flux reported in the literature
ranges from $<$ 0.5 to $1.7\times10^{-11}$ erg cm$^{-2}$
s$^{-1}$ (\cite{Halpern82}) indicating variability by around a factor of 3.
The PDS emission shows a variation of $\sim$70$\%$ during our two observations, but
due to the poor statistics in
both measurements (the signal to noise ratio in the PDS
is around $3\sigma$ level in each measurement),
we summed the observations to improve the statistics and performed an averaged
spectral analysis.
As shown in Table~\ref{tab4}, this analysis indicates a
photon index $\Gamma$ of $\sim$2.5 and an average 20--100 keV flux, corrected for
the offset, of $\sim$$8\times10^{-11}$ erg
cm$^{-2}$ s$^{-1}$. By extrapolating the flux down to the
2--10 keV range, we find a value of $\sim$$2\times10^{-10}$ erg
cm$^{-2}$ s$^{-1}$, much higher with respect to previous measurements; however,
if the photon index is restricted to values more appropriate
for a Seyfert 1 galaxy (i.e. 1.7--1.9), then the 2--10 keV flux reduces to
$\sim$$6\times10^{-11}$ erg cm$^{-2}$ s$^{-1}$, only a factor of 3 higher.
This result indicates that NGC 1566 could indeed dominate the PDS
emission and that flux variability is likely.
\begin{table*}[t]
\caption{PDS spectral analysis of serendipitous sources not detected
in the MECS instrument.}
\smallskip
\small
\label{tab4}
\begin{center}
\begin{tabular}{l c c c c c c c}
\hline
\hline
{\bf Source}& {\bf Offset$^{(a)}$} & {\bf EXPO} &{\bf cts/s}&{\bf $\Gamma$}&{\bf R$^{(b)}$}&
{\bf F$^{(c)}_{Obs}$}& {\bf F$^{(c)}_{Source}$} \\
 &  (arcmin)  & (Ks)  &  &    &   ($\%$) & [20--100] keV & [20--100] keV \\
\hline
\hline
PKS 0101--649 &32.5& 26.4 & $0.115\pm0.035$& $1.51^{+1.45}_{-1.03}$& 56 & 12.8 
& 23.0 \\
Perseus/NGC 1275 (?) & 62.0  &25.9 & $0.438\pm0.043$ & $3.27^{+0.53}_{-0.43}$& 19 &
25.0 & 136.0 \\
NGC 1566$^{(d)}$ & 60.2  &  26.6 &$ 0.192\pm0.041$& $2.50^{+1.32}_{-0.84}$& 19& 15.4
& 81.1\\
NGC 3227$^{(d)}$ & 55.1 & 176.0 & $0.247\pm0.016$ & $1.85^{+0.24}_{-0.22}$& 26 & 25.5
 & 98.1 \\
ESO 025--G002 & 31.4 &26.1 & $0.138\pm0.035$ & $2.96^{+0.91}_{-0.73}$ & 55 & 11.1
& 20.2\\
4C +74.26$^{(d)}$ &34.2  & 66.2 & $0.250\pm0.024$& $2.03^{+0.37}_{-0.31}$& 53& 24.4
& 46.0 \\
NGC 7319& 33.7 & 29.2 &$0.137\pm0.031$ & $2.07^{+0.67}_{-0.53}$&59 & 14.1 &
24.0  \\
NGC 7582 & 27.6 &56.8 &$0.490\pm0.027$ &$1.73^{+0.18}_{-0.16}$& 61 & 49.6 &
81.3\\
\hline
\hline
\end{tabular}
\begin{list}{}{}
\item[Note:]  Count rates are referred to the energy bands used in the
spectral analysis (see text).
\item[$^{(a)}$] Distance between the pointing target and the off--axis sources.
\item[$^{(b)}$] Source flux correction factor due to the PDS response to
sources not on--axis.
\item[$^{(c)}$] In units of $10^{-12}$ erg cm$^{-2}$ s$^{-1}$.
\item[$^{(d)}$] For this source the parameters of the average spectrum
are reported.
\end{list}
\end{center}
\end{table*}
\subsubsection{The AD Leonis Field}
The M--dwarf Ad Leonis has been the subject of an observational campaign
performed by \emph{Beppo}SAX (April 1997, 1, 8 and 12 May 1999).
We adopt the multi--temperature model
({\sc mekal} code in XSPEC, \cite{Mewe85}) proposed by van den Besselaar
et al. (2003)\nocite{Bess03} for the \emph{XMM--Newton} and \emph{Chandra}
observations, and 
find that the LECS and MECS data of the four observations are best
described by a two--temperature model, in which the metal abundances are
left free to vary. The 2--10 keV flux varies in the range $1.4-2.3
\times10^{-12}$ erg cm$^{-2}$ s$^{-1}$, corresponding to a
flux variation of $\sim$60$\%$, while the spectral parameters do not show
significant variation. In view of these indications, we summed together
the data from different pointings
and estimated the average spectral parameters. The two--temperature model
remains the best fit  ($\chi^{2}/\nu=97/82$), giving
$kT_{1}=0.65^{+0.03}_{-0.02}$ keV and $kT_{2}=2.04^{+0.24}_{-0.17}$ keV
metal abundances of $0.14\pm0.02$ and $0.56^{+0.25}_{-0.21}$
respectively and an averaged 2--10 keV flux of $\sim$$2\times10^{-12}$
erg cm$^{-2}$ s$^{-1}$; this spectrum if extrapolated to high energies
provides almost no contribution to the PDS flux (see Table~\ref{cont2}).

At a distance of $\sim$55$^{\prime}$ from
AD Leonis, we find the Seyfert 1 galaxy NGC 3227, which could be
the contaminating source we are searching for. NGC 3227, first detected
in the \emph{Ariel V} all sky survey, is one of
the bright sources in the Piccinotti sample of AGN (\cite{Picc82}).
NGC 3227 is known to be variable (on time scales of hours--days)
and characterized
by a spectrum with a photon index $\Gamma$ $\sim$1.6, flatter than that typically
observed in Seyfert galaxies (\cite{Georg98}).
\emph{Ginga} observations performed in 1988 indicate a 2--10 keV flux of
$\sim$$4\times10^{-11}$ erg cm$^{-2}$ s$^{-1}$ (\cite{Pounds89}), while
\emph{ASCA} observations performed during 1993 and 1995 provide 2--10 keV
flux measurements in the range $2.4-2.6\times10^{-11}$ erg cm$^{-2}$ s$^{-1}$.
Recently, Lamer et al. (2003)\nocite{Lamer03}, analyzing the \emph{Rossi X--ray
Timing Explorer (RXTE)} data of NGC 3227,
confirmed the source variability, explaining it in terms of transient
absorption by a gas cloud (neutral or weakly ionized) of column density
of $\sim$$3\times10^{23}$ cm$^{-2}$ moving across the line of sight
to the X--ray source.
This interpretation has been confirmed also by \emph{XMM--Newton} observations
(\cite{Gondoin03}).
NGC 3227 has been detected at higher energies by \emph{OSSE} and
\emph{BATSE} on--board the \emph{Compton Gamma Ray Observatory (CGRO)}.
The \emph{OSSE} data when fitted in the 50--200
keV range, show a spectral index $\Gamma=1.86^{+0.36}_{-0.38}$,
softer with respect to that found in the 2--10 keV band and
a photon flux of ($3.40\pm0.55)\times10^{-4}$ cm$^{-2}$ s$^{-1}$
in the 50--150 keV band (\cite{Zdz00}).
Based on these measurements,
we estimate a 20--100 keV flux measurement in the range
$5.0-7.8\times10^{-11}$ erg cm$^{-2}$ s$^{-1}$.
In addition, the \emph{BATSE} data (\cite{Mal99}) show
a 20--100 keV flux in the range $9.7-12.8\times10^{-11}$ erg cm$^{-2}$
s$^{-1}$.
The spectral analysis of our PDS data indicate that, during the four
observations of AD Leonis, the 20--100 keV flux measurements
of NGC 3227 ranged in the interval $6.3-14.7\times10^{-11}$ erg
cm$^{-2}$ s$^{-1}$, with an averaged value of $9.8\times10^{-11}$ erg
cm$^{-2}$ s$^{-1}$ and a photon index $\Gamma$ $\sim$1.85 (see Table~\ref{tab4}).
These results turn out to be in perfect agreement with both \emph{OSSE} and
\emph{BATSE} data.
We want to stress that our measurement contains the only spectral
information about NGC 3227 obtained by \emph{Beppo}SAX (the
source has never been observed by the satellite) and so represent an interesting
addition to the list of Piccinotti sample sources for which \emph{Beppo}SAX high energy
data are available.
\subsubsection{The H1846--786 Field}
H1846--786 is a Seyfert 1 galaxy belonging to the Piccinotti sample
(\cite{Picc82}). Despite its X--ray brightness, the source
has been poorly studied. A simple power law, with a photon index $\Gamma$
$\sim$1.95 is the best--fit
model to the LECS and MECS data ($\chi^{2}/\nu=125/100$) as obtained by
Quadrelli et al. (2003)\nocite{Quad03}.
This result indicates that no extra
absorption and/or an iron line are present in the spectrum.
The extrapolation of this best--fit model to the PDS energy range
shows a significant contribution from H1846--786 (see Table~\ref{cont2}).
In fact, as can be seen in Table~\ref{tab1}, this case represents the lowest
range for the PDS/MECS normalization constant, indicating that both the target
(H1846--786) and another object, located within the PDS
field of view, contribute to the PDS flux.
This excess emission could be due to the near ($z=0.0285$, B $ =8.8$,
R $ =10.4$)
bright Seyfert 1 galaxy ESO 025--G002, which has a \emph{ROSAT} flux
only 5 times lower than the target source.
Also this source belongs to the 2 Micron All Sky Survey eXtended Source Catalog
(i.e. 2MASX J18544039--7853544), with magnitudes J $ =11.4$, H $ =10.6$ and K $ =10.3$,
and shows weak radio emission ($\sim$10 mJy at 36 cm).
The spectral analysis of the PDS data yields a contribution
from this object of $\sim$60$\%$.
The spectrum is steep ($\Gamma$ $\sim$3) but still compatible with the canonical
AGN value of 2 and the 20--100 keV flux
turns out to be of the order of $2.0\times10^{-11}$ erg cm$^{-2}$ s$^{-1}$
(see Table~\ref{tab4} for more details).
Thus, ESO 025--G002 shows at high energies a flux of the same order of
that of H1846--786, while at soft energies is 5 time lower. This can
be explained if the source is absorbed at soft energies; in fact, fixing the PDS power law index 
to a value more appropriate to an AGN and allowing for absorption, we find
an upper limit to the column density of a few $10^{22}$ cm$^{-2}$.

Finally, we can conclude that in this particular case both target and
serendipitous source are likely to give roughly the same contribution
to the PDS emission.
\subsubsection{The VW Cephei Field}
VW Cephei, a W UMa--type binary system, has been observed by
\emph{Beppo}SAX twice (May 1998 and October 1998). Following
the investigations performed by Gondoin (2004)\nocite{Gondoin04}, we fit 
separately the LECS and MECS data of both data sets  with the 
{\sc mekal} optically thin plasma model.
In each observation, a {three--component model}, having
different temperatures but the same metal abundance, gives the best
representation of the data. The 2--10 keV flux turns out to be
$1.6\times10^{-12}$ erg cm$^{-2}$ s$^{-1}$ and 
$0.8\times10^{-12}$ erg cm$^{-2}$ s$^{-1}$ in the first and second 
pointings, respectively. Although the flux varies  
($\sim$50$\%$) from the first to the second observation,
the spectral shape shows, within the uncertainties, no evidence of
variability; therefore, we summed the two data sets together.
The three--temperature
model applied to the combined data still provides a good fit 
($\chi^{2}/\nu=37/65$) and gives temperature values of 
2.4, 0.69 and 0.13 keV; extrapolation of this model to the PDS band indicates that 
the VW Cephei contribution to the
PDS data is negligible (see Table~\ref{cont2}).

As a possible contaminating source, we find at around
$34^{\prime}$ from VW Cephei an object identified with 4C +74.26,
a radio--loud active galaxy. This is a particularly interesting
object because its \emph{ASCA} X--ray spectrum (\cite{Brink98};
\cite{Sambru99}) shows features that are
typical of Seyfert galaxies more than of Broad Line Radio Galaxies: a warm
absorber and a significant Compton reflection hump. It is also the only 
quasar in the collection of Sambruna et al. (1999)\nocite{Sambru99} with a detectable 
Fe$_{K\alpha}$ line. 
4C +74.26 was targetted by \emph{Beppo}SAX on May 1999. Analysis of these data by Hasenkopf 
et al. (2002)\nocite{Hasen02} indicated as the best--fit model 
a power law continuum modified
by Compton reflection at high energies and by absorption at low energies.
An iron line was also detected, but its energy could not be tightly
constrained, falling between 6.4 and 6.9 keV.
Our spectral analysis of the average PDS emission obtained by combining the two
observations of VW Cephei yields a 
photon index $\Gamma$ $\sim$2 and a 20--100 keV flux, corrected for
the off--axis effect, of $4.6\times10^{-11}$ erg cm$^{-2}$ s$^{-1}$ 
(see Table~\ref{tab4}).
If we fit the PDS spectrum with the {\sc pexrav} model, fixing the 
inclination angle to the value indicated by Hasenkopf et al. 
(2002)\nocite{Hasen02}, we find a  
satisfactory fit ($\chi^{2}/\nu=1.2/2$), but the reflection coefficient of 
$R$ $\sim$1.3 and the cut--off energy $E_{C}>148$ keV are not well
constrained, although in agreement with what  
reported by Hasenkopf et al. (2002)\nocite{Hasen02}.
In order to check the self--consistency of our spectral analysis, 
we reanalysed the PDS data of the May 1999 observation with a simple power 
law and obtained spectral parameters for this epoch (a photon index $\Gamma$ $\sim$2 and 
a 20--100 keV flux
of $4.0\times10^{-11}$ erg cm$^{-2}$ s$^{-1}$) consistent with those obtained during 
the VW Cephei pointings.

Overall, the data indicate that 4C +74.26 is the most likely contaminating source present in
the PDS field of view of VW Cephei and further suggest no variability in the high energy flux.
\subsubsection{The NGC 7331 Field}
A good fit to the MECS and LECS data of the LINER NGC 7331
is provided by a single power law having  a photon index
$\Gamma=1.95^{+0.12}_{-0.18}$ and 2--10 keV flux of $5.1\times10^{-13}$ erg cm$^{-2}$
s$^{-1}$.
This result confirms the non--thermal
X--ray spectrum of NGC 7331 reported by
Stockdale et al. (1998)\nocite{Stock98} on the basis of both \emph{ROSAT}
and radio observations.
The above model however gives a small contribution (see Table~\ref{cont2}) to the PDS 
emission when extrapolated to high energies.

Searching for potential high energy emitters in the HEASARC archives, we find a source 
located at about $30^{\prime}$
from the centre of the MECS pointing and identified with
the Seyfert 2 galaxy NGC 7319, which belongs to the Stephan's Quintet, a
compact group of galaxies.
Recent \emph{Chandra} data of NGC 7319 (\cite{Trinc03}) describe the
source emission as due to the superposition of a strong
and heavy absorbed nuclear source plus diffuse softer emission.
Modeling the nuclear source with a combination of
absorbed and unabsorbed power law components having the same photon index,
plus a narrow 6.4 keV emission line, the above authors found a photon index
$\Gamma$ $\sim$1.7, a column density $N_{\rm H}=4\times 10^{23}$ cm$^{-2}$,
an $EW$ $\sim$110 eV and a 2--10 keV flux of $8.2\times10^{-12}$ erg cm$^{-2}$ s$^{-1}$.
The spectral fit was similar to that derived from a previous
\emph{ASCA} observation (\cite{Awaki97}),
except for the smaller $EW$ and higher 2--10 keV flux.
The X--ray spectral parameters as well as the $F_{\rm X}$/$F_{\rm OIII}$ ratio
(\cite{Bassani99}) indicate that the source is likely to be Compton thin.
NGC 7319 ($z=0.0225$, B $ =14$, R $ =12.7$)
belongs to the 2 Micron All Sky Survey eXtended Source
Catalog (2MASX J22360355+3358327), with magnitudes J $ =11.1$, H $ =10.3$ and K $ =10.1$,
and is also detected in radio ($\sim$7 mJy at 6 cm).

The PDS data are well described by a power
law with a photon index $\Gamma$ $\sim$2.1 (see Table~\ref{tab4}) and a 20--100
keV flux of $2.4\times10^{-11}$ erg cm$^{-2}$ s$^{-1}$ at the source. By extrapolating
the PDS spectrum to the 2--10 keV band, we find a flux of
$1.7\times10^{-11}$ erg cm$^{-2}$ s$^{-1}$, which is higher by a factor of
$\sim$9 and $\sim$2 with respect to the \emph{ASCA} and \emph{Chandra}
measurements, respectively.
However, if we fit the PDS spectrum with an absorbed power law, fixing the column density
$N_{\rm H}$ and the photon index $\Gamma$ to the best--fit values found by
Trinchieri et al. (2003)\nocite{Trinc03} and extrapolate to the 2--10 keV energy
range, we find a 2--10 keV flux of
$\sim$$7\times10^{-12}$ erg cm$^{-2}$ s$^{-1}$ slightly lower than the \emph{Chandra}
value, but still compatible with it.
This result strongly indicates that NGC 7319 is a good
candidate to explain the PDS emission in the NGC 7331 observation.
\subsubsection{The NGC 7552 Field}
In this field, although two extra sources
are present within the MECS image (see Fig. 9), neither seems to be responsible
for the high energy emission seen in the PDS.

The LECS and MECS data of the starburst galaxy NGC 7552 are well
described by a two--component model, consisting of a
thermal part ({\sc Raymond--Smith} model in XSPEC)
with $kT=0.92^{+0.28}_{-0.34}$ keV and a power law with a
steep spectrum ($\Gamma=2.74^{+0.91}_{-0.75}$). The 2--10 keV
flux is $1.6\times10^{-13}$ erg cm$^{-2}$ s$^{-1}$,
and the extrapolation of this model to high energies shows an almost null
($<$ 1$\%$) contribution to the PDS emission.

As displayed in Fig. 9, within the MECS field of view we find two
other sources. The first, located at $\alpha(2000)=
23^{h}17^{m}28^{s}.8$
and $\delta (2000) = -42^{\circ} 47^{\prime} 36^{\prime\prime}.2$
is associated to the Seyfert 1 galaxy FRL 1041, while the 
second at $\alpha(2000)=
23^{h}14^{m}00^{s}.7$ and
$\delta (2000) = -42^{\circ} 43^{\prime} 17^{\prime\prime}.3$ is identified with
the cluster of galaxies Sersic 159--03.
Fitting the MECS data of FRL 1041 with
a simple power law, provides a satisfactory fit
($\chi^{2}/\nu=14.9/15$), a photon index
$\Gamma=1.63^{+0.67}_{-0.60}$, compatible with values found for Seyfert galaxies,
and a 2--10 keV flux of $5.0\times10^{-13}$ erg cm$^{-2}$ s$^{-1}$.
The extrapolation of this model to high energies gives a low (see 
Table~\ref{cont2}) contribution
and a joint fit of the MECS/PDS data gives a cross--calibration
constant (37--82) well outside the nominal range.
Concerning the galaxy cluster,
Sersic 159--03, we find that the best--fit model ($\chi^{2}/\nu=26.4/32$),
for the MECS data only, is a thermal
bremsstrahlung having a temperature value ($kT=2.54^{+0.24}_{-0.22}$ keV), compatible with
that found by \emph{XMM--Newton} observations (\cite{Kaastra01}),
plus a narrow gaussian line centered at $6.47^{+0.24}_{-0.21}$ keV
with an $EW$ of $552^{+229}_{-327}$ eV.
%
%
%
%
%
%
\begin{figure}[t]
\psfig{file=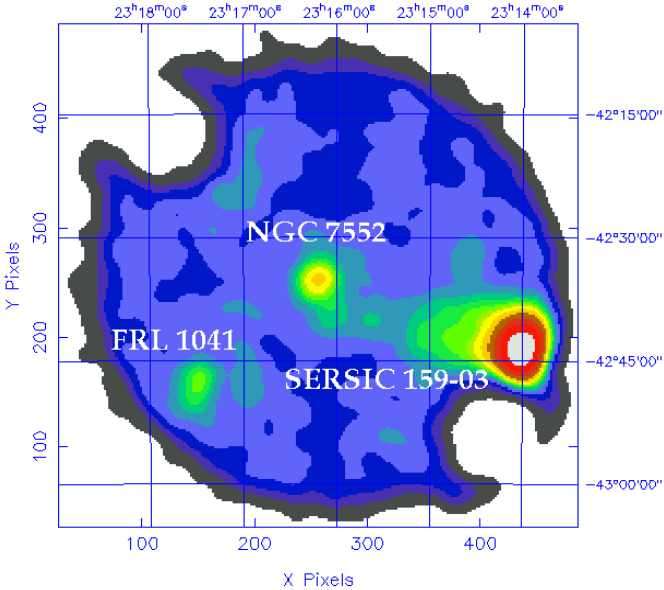,width=9.2cm,height=8cm,angle=0}
\label{campo7552}
\caption{MECS image of the sky region surrounding NGC 7552.}
\end{figure}
The 2--10 keV flux is $7.0\times10^{-12}$ erg cm$^{-2}$ s$^{-1}$.
The quality of the fit is rather worse when we fit simultaneously
the MECS and PDS data; in fact, we cannot put any constraint on the spectral
parameters and, in addition, the PDS/MECS cross--calibration
constant varies within the interval 91--303. The above results indicate
that neither source is likely to be responsible for the high
energy emission (see also Table~\ref{cont2}),
even in the case of the cluster which is strongly detected in the MECS
instrument ($\sim$50$\sigma$ level), but has an extremely soft spectrum.

In view of these findings, we performed a further search in order
to understand if there was any other nearby object which could produce the PDS 
result. Just outside the MECS field of view
we find NGC 7582, a Seyfert 2 galaxy, belonging to the Piccinotti sample,
i.e. a very bright source in hard X--rays (\cite{Picc82}; \cite{Mal99}).

NGC 7582 has already been observed by \emph{Beppo}SAX in November 1998 (\cite{Turner00}).
In that case the broad spectrum (2--100 keV) was described by a power law
of photon index $\Gamma=1.95^{+0.09}_{-0.18}$ (steeper than the index found during
a previous
\emph{ASCA} observation) transmitted through a dual absorber with column densities
$N_{\rm H}$ $\sim$$1.6\times10^{24}$ cm$^{-2}$ (covering 60$\%$ of the nucleus) and
$\sim$$1.44\times10^{23}$ cm$^{-2}$ (fully covering the source).
The authors reported a 10--100 keV flux of $\sim$1.2$\times10^{-10}$
erg cm$^{-2}$ s$^{-1}$ and the presence of variability on
time--scales down to a few thousand of seconds. In addition, Turner et al. 
(2000)\nocite{Turner00} found that NGC 7582 was significantly
brighter than the average level sampled by \emph{OSSE}
($\sim$$4\times10^{-11}$ erg cm$^{-2}$ s$^{-1}$ in the 50--150 keV band)
and \emph{BATSE} ($\sim$$8.9\times10^{-11}$ erg cm$^{-2}$ s$^{-1}$ in the 20--100 keV band)
(\cite{John97}; \cite{Mal99}), both estimated using a power law model.

By subtracting from the PDS the contribution due to each of the three sources present in the 
MECS image of 
Fig. 9 and fitting the remaining emission
with a simple power law, we find a photon index of 1.72 (see Table~\ref{tab4})
and a 20--100 keV flux (50--150 keV)
at the source of $8.13\times10^{-11}$ erg cm$^{-2}$ s$^{-1}$
($\sim$$5\times10^{-11}$ erg cm$^{-2}$ s$^{-1}$), which is in perfect
agreement with the values reported by \emph{BATSE} and \emph{OSSE}.
If we fit our data, taking into account the double absorber proposed
by Turner et al. (2000)\nocite{Turner00}, we find a photon index $\Gamma$ $\sim$1.9
and a 10--100 keV flux of $2.3\times10^{-10}$ erg cm$^{-2}$ s$^{-1}$, a factor of 2 higher
than the value reported for the 1998 observation by Turner et al. (2000).

To conclude, NGC 7582 is very likely the serendipitous source
responsible for the emission detected above 10 keV in the pointing of NGC 7552.
Furthermore, we provide evidence for variability ($\sim$$90\%$) in the flux
between the two \emph{Beppo}SAX measurements taken one year apart.
\section{Conclusions}
In this work we have shown how a careful search in the \emph{Beppo}SAX public 
archive can provide evidence for new hard X--ray 
emitting sources, most likely associated with active galaxies, 
and/or confirm the high energy emission of known objects.
In particular, we report the detection of six new hard X--ray emitters
(two type 1 Seyfert galaxies, two type 2, one quasar 
and a galaxy not yet classified as active); 
for two of these sources broad band spectra are presented,
while in four cases only results above 10 keV are reported.

For the remaining 6 objects in our sample, emission above 10 keV was known before 
this work, mainly from previous dedicated \emph{Beppo}SAX pointings.
In the case of NGC 3227, an object known to emit at high energies from past 
\emph{OSSE} and \emph{BATSE} measurements, a spectrum is published here for the 
first time.
Comparison of the present data with previous observations either with
\emph{Beppo}SAX and/or with other high energy missions indicates that flux
variability is present in three 
(MKN 766, NGC 7582 and NGC 3227) possibly four (NGC 1275) objects. 

As the AGN reported in this paper are ``loosely''
representative of the population
of extragalactic objects with emission above 10 keV, it is worth examining their
overall characteristics. 
The first consideration to make is that most of our objects (6 out of 10 with optical classification) 
are broad line emitting or unabsorbed AGN, contrary to the expectation of finding more absorbed than 
unabsorbed objects. 
The other four (possibly five if we include PKS 0101--649) show intrinsic absorption 
which, however, is compatible with a Compton thin nature ($N_{H} < 10^{24}$ cm$^{-2}$).
Only in the case of PKS 0101--649 could the source be Compton thick.
This is probably due to the limited
capability of current hard X--ray detectors which allow just the 
brightest and nearest extragalactic objects to be found. 
In fact, all but two (Perseus and 4C +74.26) or even one (if NGC 1275 is responsible for the
emission detected by the PDS) of our AGNs are nearby ($z<0.1$) and belong to
the 2 Micron All Sky Survey eXtended Source Catalog (i.e. the surrounding galaxy is visible
in the near--infrared). 
At least eight (possibly nine if we also consider NGC 1275) of our objects are radio loud,
according to Terashima $\&$ Wilson (2003)\nocite{Terashima03} definition.

Overall, we can conclude that our objects are bright, nearby, with a detectable emission
in all wavebands and only a fraction ($\sim 30-40\%$) have a column density 
in excess of $10^{22}$ cm$^{-2}$.

As shown here it is also possible to encounter offset fields
contaminated by the presence of serendipitous sources. We are now performing
a systematic analysis of all PDS data present in the archive 
(above $15^{\circ}$ in galactic latitudes) evaluating the relative offset 
fields; the aim of this work is to provide 
a list of positive detections in  these background 
measurements. We are confident that this analysis will provide new high 
energy sources as highlighted here in three cases.
\begin{acknowledgements}
This research has made use of SAXDAS linearized and cleaned event
files of LECS and MECS produced at ASI Science Data Center; and of the
High Energy Science Archive Research Center (HEASARC), provided by NASA's
Goddard Space Flight Center. We acknowledge the financial support of the Italian
Space agency (ASI) through contract ASI/CNR I/R/073/02. 
We thank J.B. Stephen for a careful reading of the manuscript.
We also thank the referee
for useful comments and suggestions.
\end{acknowledgements}
\end{document}